# DEM simulation of soil-tool interaction under extraterrestrial environmental effects

Mingjing Jiang [a,b,c,⇑], Banglu Xi [a,b,c], Marcos Arroyo [d], Alfonso Rodriguez-Dono [d]

[a] State Key Laboratory for Disaster Reduction in Civil Engineering, Tongji University, Shanghai 200092, China
[b] Key Laboratory of Geotechnical and Underground Engineering of Ministry of Education, Tongji University, Shanghai 200092, China
[c] Department of Geotechnical Engineering, College of Civil Engineering, Tongji University, Shanghai 200092, China
[d] Department of Geotechnical Engineering and Geo-Sciences, Polytechnic University of Catalonia (UPC), Barcelona, Spain

Abstract

In contrast to terrestrial environment, the harsh lunar environment conditions include lower gravity acceleration, ultra-high vacuum and high (low) temperature in the daytime (night-time). This paper focuses on the effects of those mentioned features on soil cutting tests, a simplified excavation test, to reduce the risk of lunar excavation missions. Soil behavior and blade performance were analyzed under different environmental conditions. The results show that: (1) the cutting resistance and the energy consumption increase linearly with the gravity. The bending moment has a bigger increasing rate in low gravity fields due to a decreasing moment arm; (2) the cutting resistance, energy consumption and bending moment increase significantly because of the raised soil strength on the lunar environment, especially in low gravity fields. Under the lunar environment, the proportions of cutting resistance, bending moment and energy consumption due to the effect of the van der Waals forces are significant. Thus, they should be taken into consideration when planning excavations on the Moon. Therefore, considering that the maximum frictional force between the excavator and the lunar surface is proportional to the gravity acceleration, the same excavator that works efficiently on the Earth may not be able to work properly on the Moon.

## 1. Introduction

The harsh lunar environment conditions include a low-gravity field, a ultra-high vacuum and high (low) temperatures during the daytime (night-time), which present great difference from those on the Earth (Heiken et al., 1991; Ouyang, 2005). In recent decades, many analytical techniques (Godwin and Spoor, 1977; Patel and Prajapati, 2011; Zeng et al., 2007; McKyes and Ali, 1977; Hettiaratchi and Reece, 1967; Blouin et al., 2001; Kuczewski and Piotrowska, 1998; Willman and Boles, 1995; Grisso and Perumpral, 1985; Kobayashi et al., 2006) of soil-tool interaction have been developed for analyzing lunar excavation on the Earth, and extensive experimental tests (Kobayashi et al., 2006; Boles et al., 1997; Boles and Connolly, 1996; Green et al., 2012; Onwualu and Watts, 1998; King et al., 2011; King and Brewer, 2012; Johnson and King, 2010; Iai and Gertsch, 2012; Agui et al., 2012) have been conducted with lunar regolith simulant. Boles et al. (1997) and Boles and Connolly (1996) investigated the gravity effects by conducting tests on an aircraft performing appropriate parabolic flight maneuvers. Their results showed that the gravity term is not directly related with the excavation force. Further

⇑ Corresponding author at: State Key Laboratory for Disaster Reduction in Civil Engineering, Tongji University, Shanghai 200092, China.
 E-mail address: mingjing.jiang@tongji.edu.cn (M. Jiang).

researches on gravity effects are necessary to verify the availability of those analytical techniques on the lunar environment to reduce the risk of lunar excavation missions. However, in spite of the high cost, experimental tests on the Earth cannot reproduce other extraterrestrial conditions (i.e., ultra-high vacuum), which may alter the lunar regolith properties and therefore affect the soil-tool interaction.

The ultra-high vacuum condition leads to a very thin layer of adsorbed gas on the lunar regolith surface. According to Perko et al. (2001), considerable inter-molecular forces, including van der Waals forces and electrostatic forces, are induced when the intermolecular distance between the molecules of two contacting particles is in the order of their molecule size. It is pointed out that the electrostatic forces are negligible for lunar grains and only van der Waals forces need to be considered. Experimental data obtained by Bromwell (1966) and Nelson (1967) under ultrahigh vacuum ($10^{-7}$ Pa) and high temperature (394 K) indicated an increase friction angle of 13º and an increase cohesion of 1.1 kPa testing simulated lunar soil. Increased cohesion of lunar simulant in high vacuum was also reported by Johnson et al. (1973, 1970).

In such circumstances, numerical simulation seems to be a useful method in studying soil-tool interaction under extraterrestrial environments, because it can provide some information useful to assess the applicability of theoretical models and to correctly design the most suitable type of excavator for such environments. Some researchers employed the Finite Element Method to study soil-tool interaction by predefining the failure surfaces (Abo-Elnor et al., 2003; Abo-Elnor et al., 2004; Zhong et al., 2010). On the other hand, the Distinct Element Method (DEM) Cundall and Strack, 1979 is a numerical technique which plays a very important role in granular material research, especially in soil-tool interaction where large soil deformations and failure appear. Although the grains and the inter-particle contact are modeled in an idealized way, this method can still provide a useful insight into the mechanical behavior (Mak et al., 2012; Chen et al., 2013; Coetzee and Els, 2009; Asaf et al., 2007; Obermayr et al., 2011; Coetzee and Els, 2009; Shmulevich, 2010; Shmulevich et al., 2007; Nakashima et al., 2011; Nakashima et al., 2008; Bui et al., 2009). However, most researches focused on the mechanism of soil-tool interaction applied to the Earth (Mak et al., 2012; Chen et al., 2013; Coetzee and Els, 2009; Asaf et al., 2007; Obermayr et al., 2011; Coetzee and Els, 2009; Shmulevich, 2010; Shmulevich et al., 2007; Nakashima et al., 2011) and the others lacked a suitable contact model fully taking into account the features of the lunar regolith grains and considering the unique lunar environment, both of which affect the soil-tool interaction significantly. For instance, Chen et al. (2013) used three-dimensional (3D) simulation to simulate a slurry injection tool and its interaction with soil in three different soil fields (coarse sand, loamy sand and sandy sand). Obermayr et al. (2011) used DEM to predict the draft force of a simple implement in a cohesionless granular material and the results were compared with those of small-scale laboratory tests using steel balls or round washed gravel. Nakashima et al. (2008) used DEM to investigate the gravity effects on rotational cutting tests of lunar regolith simulant and compared the results with those of laboratory tests on a parabolic pattern of flight of an airplane. But they used the linear contact model with constant spring stiffness. Bui et al. (2009) introduced an interlocking force to the conventional normal force to characterize the lunar soil properties in order to study the gravity effects. But they focused on the gravity effects on the ultimate bearing capacity of lunar regolith and lacked enough analysis on soil cutting test.

In view of the unsatisfactory contact models employed in DEM simulations of soil cutting test so far, Jiang et al. developed a novel contact model of lunar regolith taking inter-particle rolling resistance and van der Waals forces into consideration (Jiang et al., 2013). The model can capture the mechanical behavior of lunar regolith (high peak friction angle and apparent cohesion) and enables highly efficient DEM simulation of lugged wheel performance (Jiang et al., 2014a, 2014b) on the lunar environment. In this paper soil cutting tests were numerically simulated to capture the unique features of soil-tool interaction under lunar conditions, including the effects of the lunar gravity field and the van der Waals forces. The mechanism of the effects is also analyzed by observing the soil response, which is difficult to observe in experimental tests but extremely important in developing the available analytical model. Some other features that may also affect the soil-tool interaction (e.g. the geometry of the blade) are not presented here. Compared with the Reference Jiang et al. (2015a), where only the gravity effects are briefly introduced in Chinese, this paper is a comprehensive and complete one. The present DEM study was performed in 2D, because a 3D contact model is still being developed by the authors and is less affordable in terms of computational time and particle number for a large boundary-value problem. Although 2D simulations have limits in the realistic representation of the soil grains and in the volumetric response, it can still provide some insight into the fundamental mechanisms of soil-tool interaction under extraterrestrial condition. It is one of our future works to carry out 3-D DEM simulation to study the effects of the lunar environment to obtain more accurate results with latest workstations.

2. DEM model of lunar regolith

Compared with the sand on the earth, the lunar regolith shows a higher friction angle and an apparent cohesion. These features are mainly caused by two reasons: the inter-particle rolling resistance due to the particle shape and the van der walls force due the high vacuum conditions (Jiang et al., 2013). Here only some brief introduction is provided. More details can be obtained in the reference (Jiang et al., 2013).

### 2.1. Inter-particle rolling resistances

Lunar regolith grains are mainly composed of angular/sub-angular agglutinates and breccias with rough surface and relatively smooth spherical glasses, while those used in DEM simulation are idealized as disks (2D) or spheres (3D). However, inter-particle rolling resistance plays a role between rough and irregular grains and thus it should be taken into consideration. Jiang et al. (2013) formulate the rolling resistance at the contact as shown in Eq. (1)

$$M = \begin{cases} \frac{K_n \beta^2 r^2}{12} \theta, & M < \frac{K_n u_n \beta r}{6} \\ \frac{K_n u_n \beta r}{6}, & M \geq \frac{K_n u_n \beta r}{6} \end{cases} \quad (1)$$

where $K_n$ is the normal contact stiffness and $u_n$ is the overlap, $h$ is the relative rotation angle and $b$ is a parameter introduced here as a shape parameter to link the contact size with the average particles radius $r$. Jiang et al. (2013) have shown that a high friction angle can be attained in DEM simulations by introducing a rolling resistance into the contact model, which features the actual mechanical behavior of lunar regolith (Jiang et al., 2013).

### 2.2. Van der Waals forces

Lunar environmental conditions include ultra-high vacuum and high temperatures, which lead to the creation of a thin layer of gas molecules adsorbed on the particles surface. Thus the van der Waals forces need to be considered on the Moon while this is not necessary on the Earth. Jiang et al. (2013) formulate the van der Waals forces between two smooth flats with infinite depth as in Eq. (2).

$$F_v = \frac{A \beta r}{6 \pi D^3} \quad \text{(Flat - flat, per unit area)} \quad (2)$$

where $A$ is the Hamaker constant, $4.3 \times 10^{-20}$ J here (Perko et al., 2001), and $D$ is the thickness of the adsorbed gas molecules.

The contact laws are illustrated in Fig. 1. Fig. 1a shows that the normal contact law includes two branches: the linear-elastic contact force and the van der Waals forces. In the model, both branch 1 and branch 2 plays a role when two particles are in contact and they disappear when two particles are separated. Fig. 1b and c indicates that the tangential and rolling strength are controlled by the overlap ($u_n$), which increases when the van der Waals forces are included. As a result, the van der Waals forces increase the macroscopic shear strength by enhancing both tangential and rolling strengths.

In summary, two different contact models are employed in the paper, with and without van der Waals forces, representing the high-vacuum environment in extraterrestrial condition and non-vacuum environment in terrestrial condition, respectively. Then the effects of the extraterrestrial environment, including extra-high vacuum, high (low) temperatures and lower gravity, can be investigated by employing different models and gravity conditions.

### 2.3. Model calibration

The model has been calibrated through a parametric study made with the help of DEM simulations of a biaxial compression test by Jiang et al. (2013). The model parameters are selected in such a way that the macroscopic strength values fall into a reasonable range for real lunar regolith. Table 1 presents the parameters used in DEM simulations, while the particle size distribution is shown in Fig. 2. Here we did not choose a curve within the upper and lower bounds of real lunar regolith but, instead, used a curve with relatively uniform particle size, ranging from 1.5 mm to 2 mm in diameter. This is for computational reasons since the efficiency greatly reduces with the widening of the particle size distribution. Fig. 3 presents the stress-strain relationships from Jiang et al. (2013). It is shown that strong strain-softening behavior can be observed under both two environments (lunar environment and terrestrial environment) and the peak strength values under the extraterrestrial environment ($\varphi = 42.6°$, $c = 0.2$ kPa)

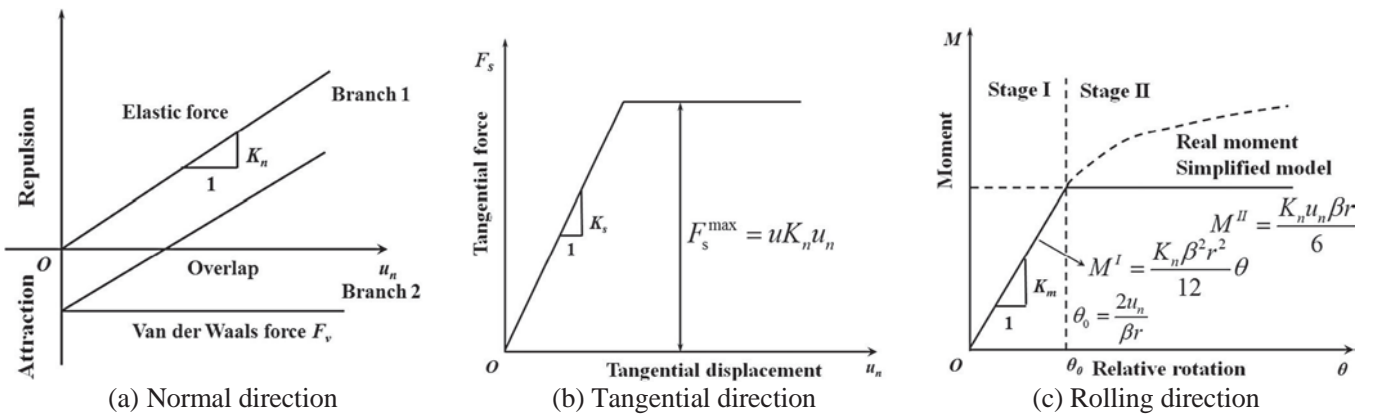

(a) Normal direction     (b) Tangential direction     (c) Rolling direction

Fig. 1. Mechanical responses of the contact law.

Table 1
Parameters for lunar soil bin in DEM simulations.

| Name | Symbol | Value |
| --- | --- | --- |
| Void ratio | $e$ | 0.22 |
| Dry density/kg/m$^3$ | $\rho$ | 2600 |
| Normal stiffness/N/m | $k_n$ | 7.5x10$^7$ |
| Tangential stiffness/N/m | $k_s$ | 5.0x10$^7$ |
| Frictional coefficient | $u$ | 1.0 |
| Rolling resistance coefficient | $\beta$ | 1.3 |
| Thickness of adsorbed molecular layer /m | $D$ | 1.5x10$^{-8}$ |

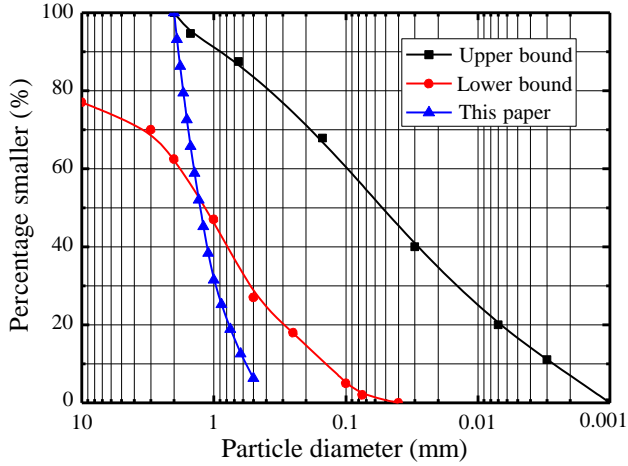

Fig. 2. Distribution of grain size used in the DEM analysis (Bui et al., 2009; Jiang et al., 2014b).

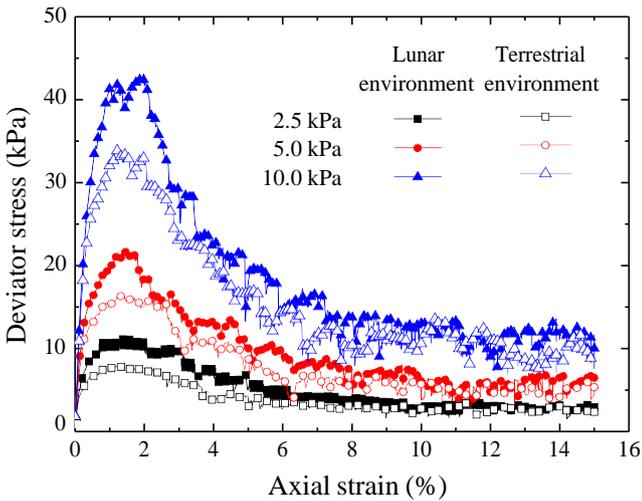

Fig. 3. Deviator stress against axial strain in lunar and terrestrial environments (Bui et al., 2009).

are larger than those under the terrestrial environment ($\varphi$ = 38.8°, c = 0 kPa).

In PFC$^{2D}$, both local and viscous damping are available. Considering that soil-tool interaction is a dynamic problem rather than a quasi-static problem in biaxial compression tests, the viscous damping is used for the DEM simulations in this paper because it is more suitable in dynamic simulations. The damping coefficients are selected by comparing the results of the simulation with the experimental results of the repose angle test, a simple dynamic problem. The relationship of damping coefficients against repose angle is presented in Fig. 4. It is shown that the repose angle increases with damping coefficient. 0.4 is selected because its corresponding repose angle is 31.25°, which falls into the reasonable range from 30° to 33° according to experimental tests of TJ-1 simulant (Jiang et al., 2015b; Dai, 2013) (a relatively novel soil simulant developed at Tongji University in Shanghai, which has been widely used in Chinese lunar-related scientific research projects (Jiang et al., 2013a, 2013b, 2013c). In addition, gravity effects on repose angle are investigated and the results are shown in Fig. 5. It can be observed that there is no significant difference on repose angle under different gravity fields, which is the same conclusion of Nakashima et al. (2011). Therefore we can choose a viscous damping coefficient equals to 0.4 for our simulation.

## 3. DEM model of soil-tool interaction

### 3.1. DEM ground

The DEM model dimensions are presented in Fig. 6. The model has a length of 0.5 m and a height of 0.15 m. 78,000 particles are compressed using the Multi-layer Under-Compaction Method proposed by Jiang et al. (2003) to ensure the homogeneity with a planar void ratio of 0.22. Then the ground is consolidated under different gravity fields (1/6 g, 1/2 g, 1 g, 2 g and 5 g) with different contact models. It is worth mentioning that the terrestrial environment corresponds to the contact model without van der Waals forces and the extraterrestrial environment corresponds to the contact model with van der Waals forces. The gravity density of the ground is $\gamma$ = 20.29 kN/m$^3$ and the coefficient of earth pressure at rest ($K_0$) is 0.509 which is obtained by dividing the horizontal stress by vertical stress as shown in Fig. 7. The distribution patterns of vertical stress and horizontal stress are the same under different gravity fields and the values are summarized in Table 2. It is shown that both the vertical stress and horizontal stress are proportional to the gravity, and different gravity fields show no significant effects on $K_0$ while the van der Waals forces make the $K_0$ lower under the high-vacuum condition.

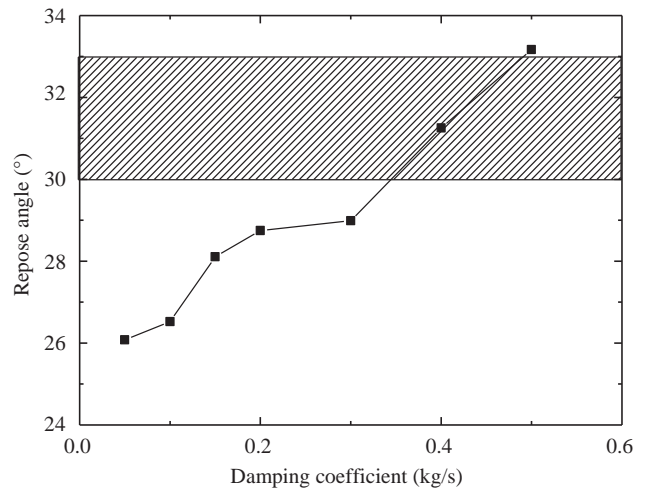

Fig. 4. Repose angles under different damping coefficients.

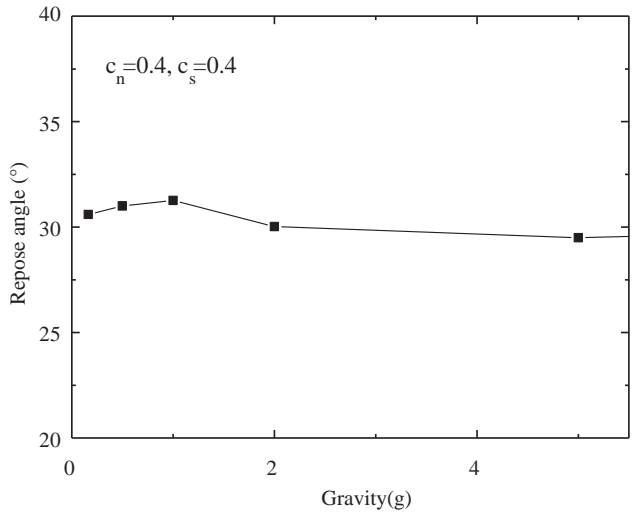

Fig. 5. Repose angles under different gravity fields.

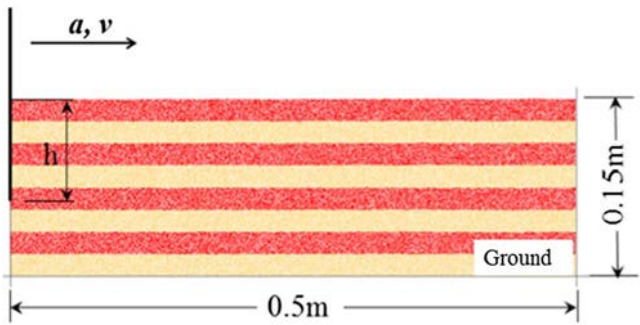

Fig. 6. The DEM model for soil cutting test.

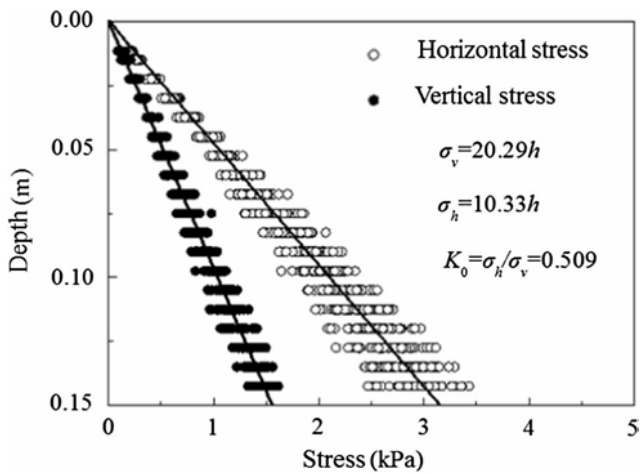

Fig. 7. The distribution of vertical and horizontal stresses under 1 g gravity field.

### 3.2. Test process

The blade is initially placed on the left side of the soil bin vertically. After an acceleration period of 0.5 s during which the cutting acceleration is $a = 2v$, the blade moves to the right side with a constant cutting speed. The cutting depth is 0.07 m, the attack angle is 0° and the cutting speed is 0.01 m/s. Two sets of simulations were conducted with different models. Each set includes five simulations under different gravity fields. It is worth mentioning that the particle packing of all the simulations is identical before consolidation; it differs during consolidation with different gravity fields and contact models.

## 4. Soil response in DEM ground

The soil response in the DEM ground is studied here from both microscopic and macroscopic points of view because it can reveal the slide surface and the affected area. The slide surface plays an important role in developing theoretical models for soil cutting tests. In laboratory experiments the slide surface is difficult to be observed directly and in FEM simulations the slide surface is always predefined before the cutting process, which seems not reasonable. However, DEM simulations have an advantage in observing the evolution of microscopic phenomena to reveal the slide surface. The total displacement field of the particles, the PDR (Particle Displacement Ratio) (Coetzee and Els, 2009) and the velocity field (Chen et al., 2013; Obermayr et al., 2011; Shmulevich et al., 2007) have been employed to try to represent the slide surface. The PDR is a ratio defined by the magnitude of the absolute displacement vector of the particles divided by the magnitude of the blade displacement. Since the velocity of the blade is constant, the PDR is identical to the displacement field at the same cutting blade displacement, so it has not been included here. In addition, the void ratio has also been employed in this paper for revealing the failure surface.

### 4.1. Soil heap structure

As the blade moves, the soil accumulates in front of the blade, which results in a surcharge which leads to an increase of the cutting resistance. Fig. 8 presents the soil heap structures on terrestrial conditions under 1 g gravity field. Fig. 8a shows the simulation results and Fig. 8b presents the experimental results of Coetzee and Els (Coetzee and Els, 2009). It is shown that the soil heap accumulates higher and bigger in front of the blade as its slope becomes steeper, till it is nearly parallel to the repose angle line, in both simulation results and experimental results. Most analytical models (Zeng et al., 2007; Kobayashi et al., 2006) assume that this surcharge is equivalent to a uniform load or triangle load distribution. According to our research a trapezium load distribution is recommended in developing analytical models.

Table 2
Initial stress condition of the ground under different gravity.

| Environment | Terrestrial environment | | | Extraterrestrial environment | | |
|---|---|---|---|---|---|---|
| Gravity field | Vertical stress | Horizontal stress | $K_0$ | Vertical stress | Horizontal stress | $K_0$ |
| 1/6 g[*] | $\sigma_v$ = 3.49 h | $\sigma_h$ = 1.76 h | 0.504 | $\sigma_v$ = 3.56 h | $\sigma_h$ = 1.76 h | 0.504 |
| 1/2 g | $\sigma_v$ = 10.46 h | $\sigma_h$ = 5.22 h | 0.499 | $\sigma_v$ = 10.76 h | $\sigma_h$ = 4.163 h | 0.387 |
| 1 g | $\sigma_v$ = 20.29 h | $\sigma_h$ = 10.33 h | 0.509 | $\sigma_v$ = 20.36 h | $\sigma_h$ = 8.11 h | 0.398 |
| 2 g | $\sigma_v$ = 41.85 h | $\sigma_h$ = 20.45 h | 0.488 | $\sigma_v$ = 41.73 h | $\sigma_h$ = 16.87 h | 0.407 |
| 5 g | $\sigma_v$ = 104.96 h | $\sigma_h$ = 50.39 h | 0.480 | $\sigma_v$ = 106.97 h | $\sigma_h$ = 44.96 h | 0.420 |

[*] $g$ = 9.8 N/kg.

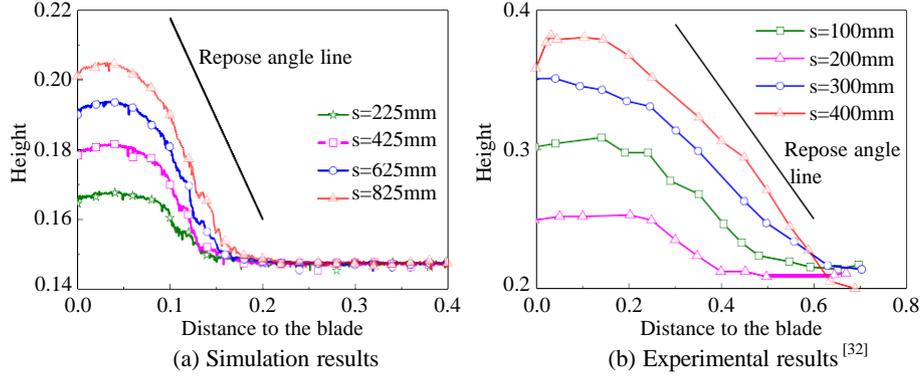

Fig. 8. Soil heap structure in simulation results and experimental results.

Fig. 9 presents the soil heaps under different gravity field on both high-vacuum and non-vacuum environments. It is shown that the soil heap evolves more precipitously in lower gravity fields. The van der Waals forces can also make the soil heap slope steeper when the gravity is low while the effects are reduced when the gravity field is high. In addition, the soil heap can be simplified as a trapezium distribution when developing analytical models as shown in Fig. 9, with the inclined boundary parallel to the repose angle line.

### 4.2. Total particle displacement field

Fig. 10 provides the particle displacement fields corresponding to 1/6 g, 1 g, and 5 g gravity fields in both high-vacuum and non-vacuum conditions. The particles are colored differently according to its displacement. It can be observed that when the gravity field decreases, the affected area becomes larger in both high-vacuum and non-vacuum conditions. Comparing the displacement fields on high-vacuum and non-vacuum conditions under the same gravity field to analyze the effects of van der Waals forces, it is obvious that the affected areas are larger on the non-vacuum condition.

### 4.3. Void ratio

In Fig. 11, maps of the distribution of void ratios in the ground are shown. It can be observed that there exists a visible affected area in front of the blade, where the void ratio is bigger than in the area far from the blade. This is because the soil in our simulation is dense and soil dilatancy appears within the affected area. Fig. 11 reveals that, the affected areas are larger on terrestrial conditions and lower gravity fields, which is consistent with the conclusion we have drawn in the analyses of the displacement field. A bold line which presents a void ratio equal to 0.21 has been drawn for comparative purposes.

### 4.4. Velocity field

A more realistic way to determine the failure surface is to observe the velocity field of the soil particles. The sliding surface would be located where the velocity gradient is higher. Normalized by the blade velocity in each case, all velocities of the particles are divided into five groups of magnitudes and rendered with different colors as shown in Fig. 12. The velocity vectors described by different colors represent the sliding line of the particles, which can reflect the failure mechanism. It can be observed that, the particle velocity decreases rapidly from 0.01 m/s to 0.02 m/s within a narrow shear band. Assuming that the particle is affected when its velocity is bigger than 0.02 m/s, the affected number of particles (No.) and their average velocity (Vel.) are summarized in Fig. 12. It can be observed that as the gravity field increases, the affected area and thus the number of particles becomes smaller, but the average velocity is larger. In addition, the affected area is smaller and the average velocity is larger under a high-vacuum environment.

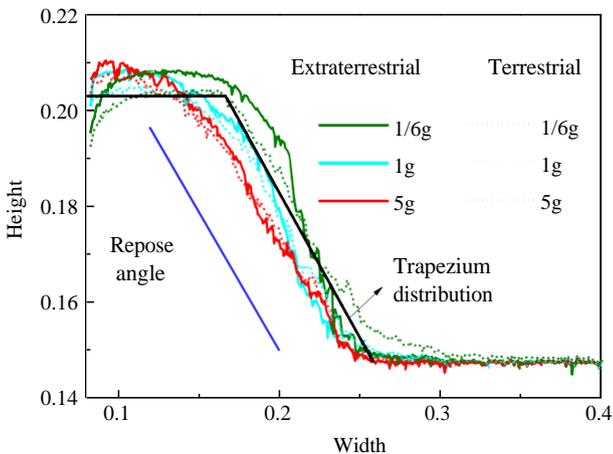

Fig. 9. Soil heap structure in terrestrial and extraterrestrial environments.

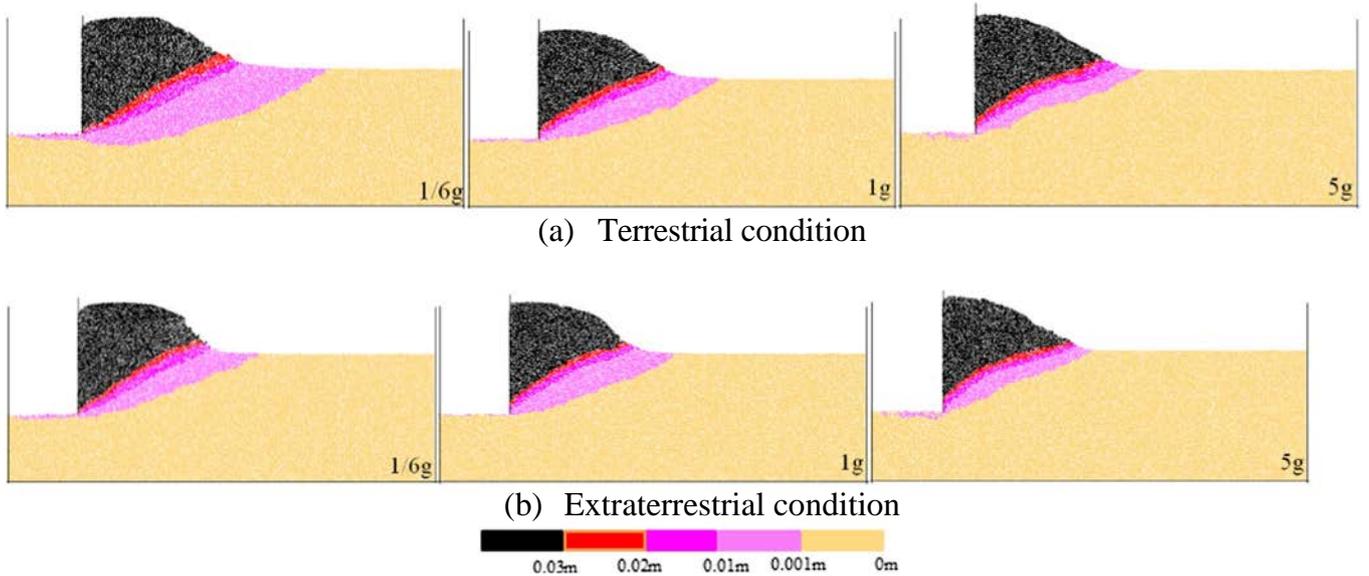

Fig. 10. Total particle displacement field under different conditions.

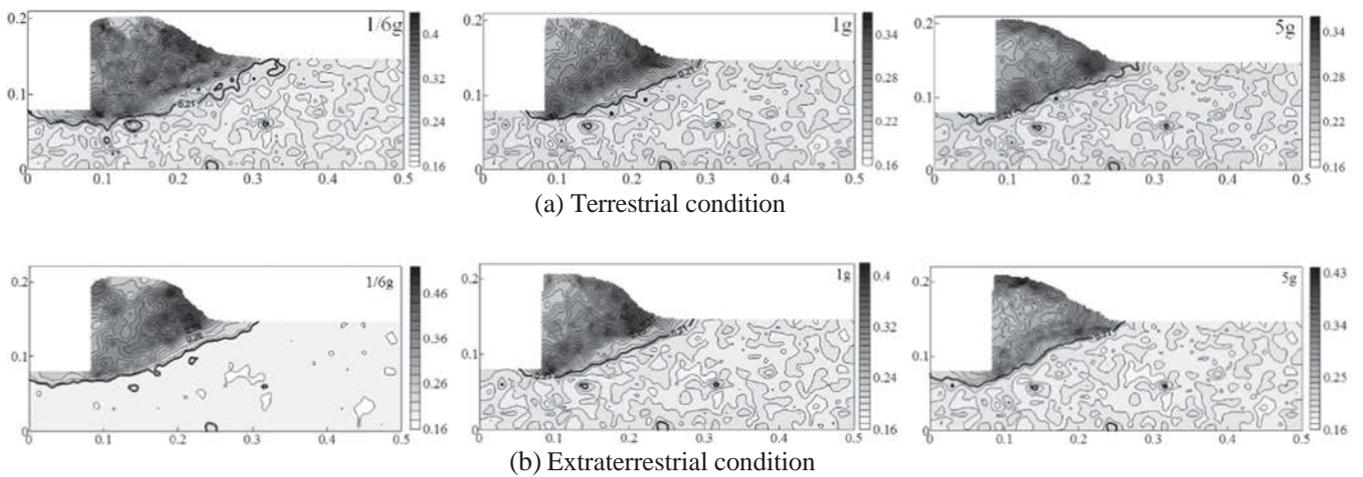

Fig. 11. The distribution of void ratio under different conditions.

As mentioned above, the harsh lunar environment conditions present significant effects on the affected area during the cutting test. The lower gravity field makes the affected area larger while the van der Waals forces make the affected area smaller. The effects of van der Waals forces can be taken into consideration by enhancing the soil strength, but the effects of gravity are not included in analytical techniques: the affected area and the slide surface are assumed to be the same under different gravity fields.

5. Blade performance

This section focuses on the effects of gravity and van der Waals forces on the performance of the blade, including the cutting resistance, the bending moment and the energy consumption. Because the cutting resistance is related to the capability of the excavator, the energy consumption is related to the weight of lunar regolith that is to be excavated and the bending moment is related with the design

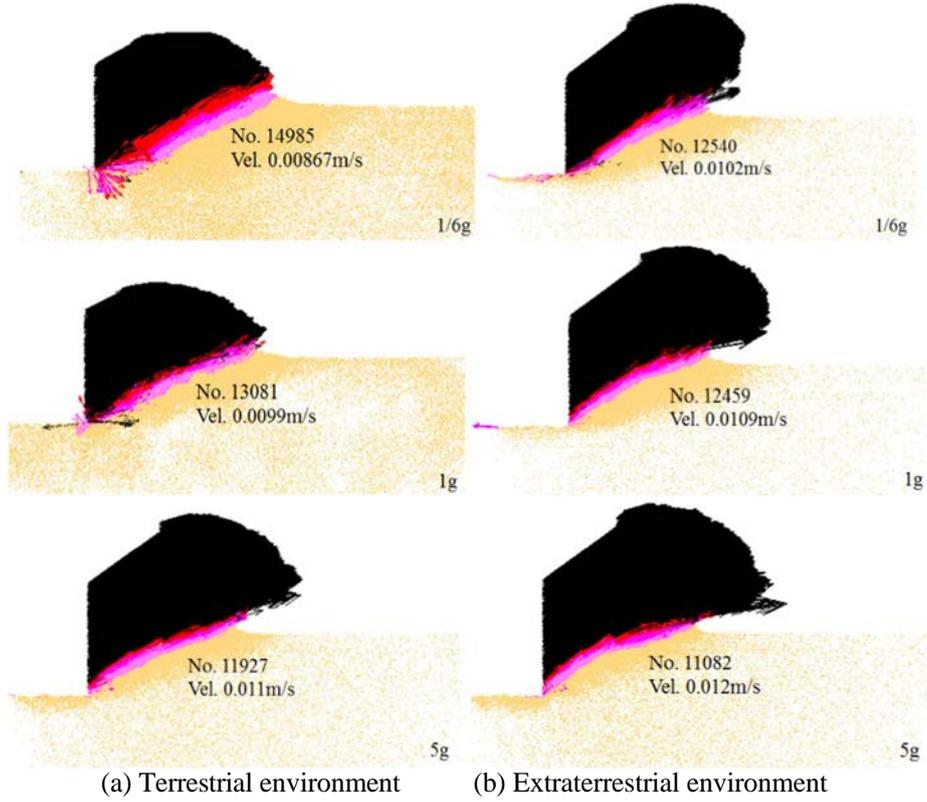

(a) Terrestrial environment     (b) Extraterrestrial environment

Fig. 12. Velocity field under different conditions (Blade velocity = 0.01 m/s).

of the bucket, all these three aspects are very important in an eventual lunar exploring plan.

In the cutting process, the cutting resistance and the bending moment are recorded during the cutting process, while the energy consumption is calculated here with Eq. (3).

$$W = \sum F \Delta s \qquad (3)$$

where $F$ is the cutting resistance and $\Delta s$ is the displacement increment.

### 5.1. Cutting resistance

In lunar excavation, the cutting resistance needs to be balanced by the frictional force between the excavator and the lunar surface, which is proportional to the weight of the excavator. In early lunar exploration, the excavators need to be sent to the Moon from the Earth, so the lighter the excavators, the more economic would be the lunar excavation mission. Also, a smaller cutting resistance means that lighter excavators are needed. Therefore, it is necessary to study the effects of gravity on cutting resistance.

Fig. 13 presents the evolutions of cutting resistances under different gravity fields on high-vacuum and non-vacuum environments. The evolution has been smoothed by moving average to show the evolution trend, considering that the raw data has large fluctuations. It can be observed that the cutting resistance increases from 0 to a peak value rapidly at the beginning of the test. After the peak value, the cutting resistance decreases to a lower value due to soil dilatancy. Then, the cutting resistance increases gradually as the blade moves on. All the curves exhibit the same evolution trend.

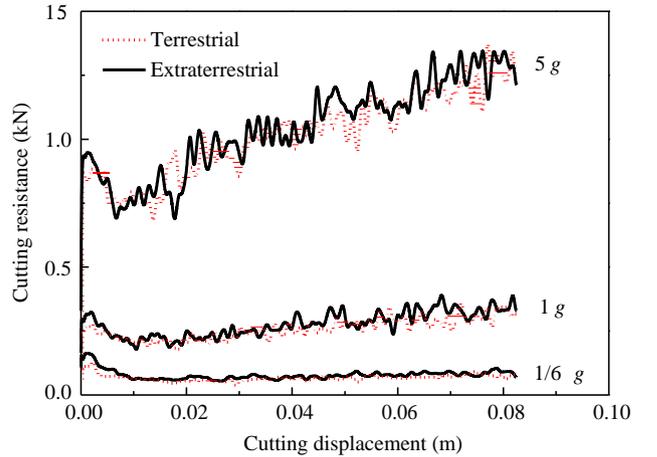

Fig. 13. Evolution of cutting resistance under different conditions.

Nevertheless, the initial peak value mainly depends on the soil properties while that gradual increase of the cutting resistance mainly depends on the weight of the accumulated soil. In fact, when the gravity is large (5 g), the weight of the accumulated soil plays a more important role than the soil properties and the initial peak value of the cutting resistance is smaller than the one achieve later on due to the accumulation of the soil heap. However, when the gravity is small (1/6 g), the effect of the soil properties becomes more important than the effect of the accumulated soil, and as a consequence the initial peak value of the cutting resistance is much bigger than the one achieve later on. Considering that the gravity field on the lunar surface is about 1/6 of that of the Earth, here we focus on studying the initial peak value of the cutting resistance, since in that case it would be the highest value achieved.

Fig. 14 provides the relationship between the initial peak value and the gravity field on both high-vacuum and non-vacuum environments. For analytical purposes, we have selected the empirical equation for the cutting resistance (P) proposed in Obermayr et al. (2011), shown in Eq. (4).

$$P = (\rho_b g z^2 K_p + c z K_c + c_a z K_{ca} + \rho_b g v^2 K_a + q z K_q) w \quad (4)$$

where $\rho_b$ is soil density, $c$ is cohesion, $c_a$ is adhesion, $z$ is cutting depth, $w$ is blade width, $v$ is cutting speed, $q$ is surcharge load, $K_p$, $K_c$, $K_{ca}$, $K_a$, $K_q$ are coefficients for earth pressure, cohesion, adhesion, velocity and surcharge load. According to the Eq. (4), if the cohesion $c$ = 0 kPa and the blade is smooth ($c_a$ = 0) under terrestrial conditions, the cutting resistance is in direct ratio with the gravity field. When the gravity is nearly 0 g, the cutting resistance is 0 N. However, as shown in Fig. 15, although the resistance forces can be fitted by a linear equation, we can observe that the cutting resistance isn't 0 N when the gravity filed is 0 g. This is because when the gravity decreases, the initial stress state also decreases, which leads to the decreasing of the cutting resistance, but the lower gravity also makes the affected area larger, as mentioned earlier regarding the soil response, which increases the cutting resistance. Under the terrestrial conditions where the apparent cohesion is 0.86 kPa, the presence of the apparent cohesion is also the reason why the cutting resistance isn't 0 N when the gravity field is 0 g. In contrast, the experimental results (Boles et al., 1997; Boles and Connolly, 1996) exhibit a similar evolution trend, but the increasing rate is different, because of the different tool type, three-dimensional effects. It is quite difficult to mimic the lunar environment on the Earth to conduct more experimental validation of our model.

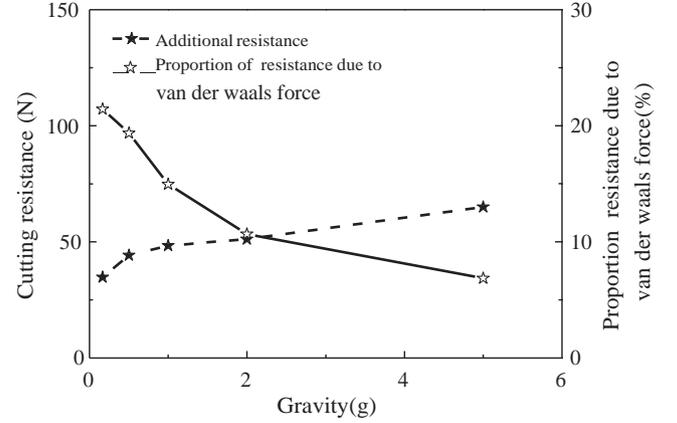

Fig. 15. Additional cutting resistance due to van der Waals forces under different gravity fields.

According to Jiang et al. (2013), the van der Waals forces can enhance the macroscopic shear strength. Thus, the cutting resistance increases significantly when the van der Waals forces are considered on extraterrestrial conditions. Fig. 15 shows the cutting resistance due to the effect of the van der Waals forces under different gravity fields. It is shown that this resistance increases with the gravity field. This is because that the coordination number increase with the gravity field as shown in Fig. 16, which shows the coordination number at the end of the test. Also we can see that the coordination number on the non-vacuum condition are larger than the on the high-vacuum condition. This is because the soil shows more significant shear dilatancy. However, the proportion of this additional resistance in the total cutting resistance decreases rapidly when the gravity field increases. When the gravity is 5 g, this proportion is 6.8% which may be ignored while when the gravity is 1/6 g it is 21.4% which should not be ignored. So in lunar excavation, the effects of van der Waals force should not be ignored.

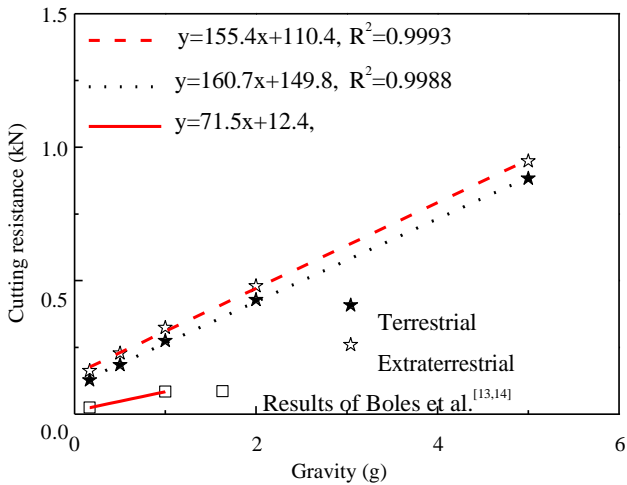

Fig. 14. Relationship between peak excavation force and gravity acceleration.

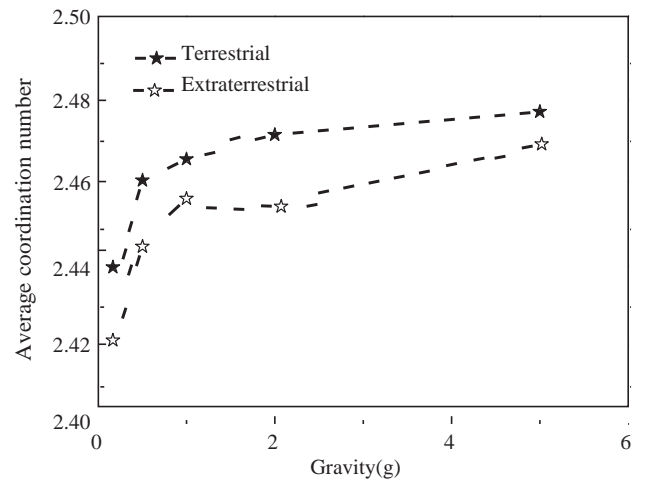

Fig. 16. Average coordination number under different conditions.

*5.2. Moment*

During the excavation process, the joint of the cutting blade endures both the effects of the cutting resistance and the bending moment, so it should be well designed to mitigate its damage. The evolution of the bending moment is shown in Fig. 17. It can be observed that the bending moment firstly increases from 0 to a peak value rapidly, at the beginning of the test. After the peak, the bending moment decrease to a lower value. Then the bending moment increase again as the blade moves on. The bending moment presents a similar developing trend as with the cutting resistance and for the same reasons that in the analysis of the cutting resistance.

The moment arm, which shows the location of the centroid of the exerted forces on the blade, can provide some perspective on this behavior. The moment arm here is calculated by dividing the bending moment with the cutting resistance. Fig. 18 presents the relationship between the moment arm and the gravity field under both high-vacuum and non-vacuum environments. The moment arm under a high-vacuum environment is a little larger than that under a non-vacuum environment, but the difference is not significant, so it can be ignored. However, the effects of the gravity field on the moment arm cannot be ignored when the gravity field is small, as shown in Fig. 17. In fact, the difference between 1/6 g and 2 g gravity fields is around 0.007 m which is 10% of the cutting depth.

Fig. 19 provides the relationship between the initial peak value of the bending moment and the gravity field on both high-vacuum and non-vacuum environments. Fig. 19 shows that the bending moment increasing rate is a bit higher when the gravity field is low. This is consistent with the moment arm increasing rate shown in Fig. 18 for low gravity fields. Fig. 20 shows that the additional bending moment, related to the additional cutting resistance and the moment arm, increase with increasing gravity field. However, the proportion of additional bending moment due to van der Waals forces decreases rapidly with increasing gravity field. When the gravity is 1/6 g, the gravity on lunar surface, this additional bending moment is 28% of the total bending moment. Therefore, we can say that the van der Waals forces should not be ignored in lunar excavation.

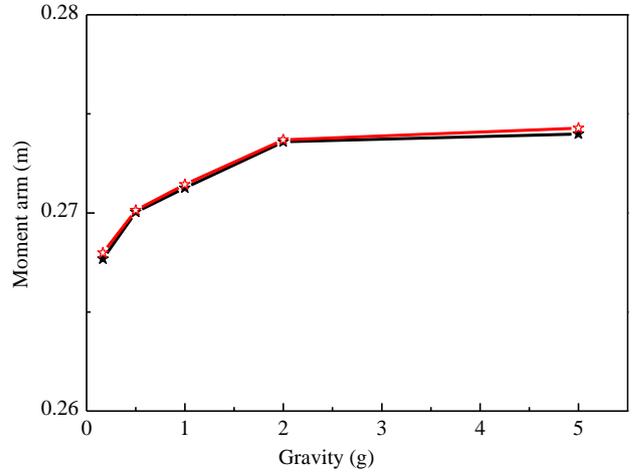

Fig. 18. Evolution of the moment arm under different conditions.

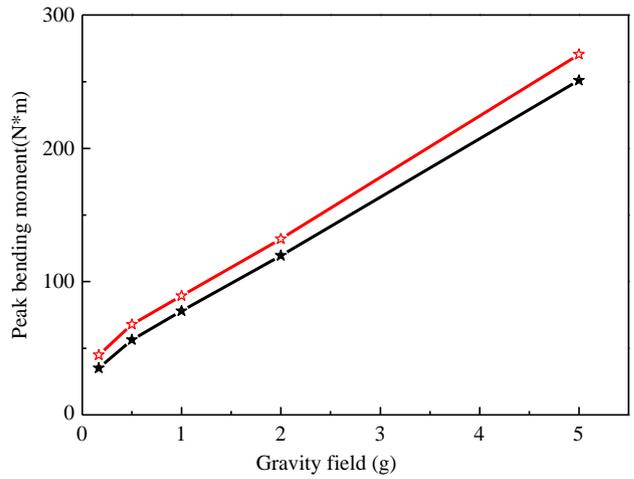

Fig. 19. Relationship between initial peak moment and gravity acceleration under different conditions.

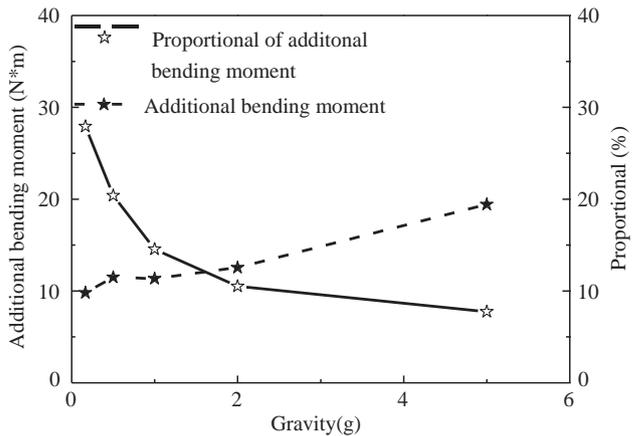

Fig. 20. Additional bending moment due to van der Waals forces under different gravity fields.

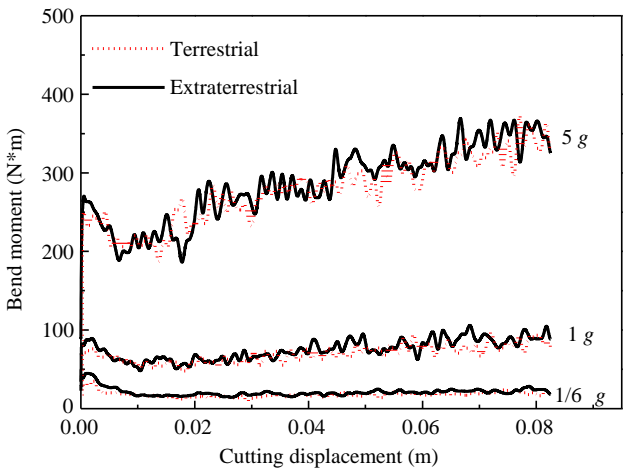

Fig. 17. Evolution of the bending moment under different conditions.

*5.3. Energy consumption*

In early lunar exploration, most of the necessary energy can be provided by the Sun. But in order to produce solar energy, we need solar panels and other equipment that have to be brought from the Earth, which is very expensive mainly due to launch costs. Therefore, in lunar excavation it is important to be energy efficient and, hence, reduce the energy consumption during the excavation as much as possible.

Fig. 21 presents the evolution of the energy consumption in the cutting process under different gravity fields and on both high-vacuum and non-vacuum conditions. It can be observed that the energy consumption evolves similarly in two environments, so it seems that the van der Waals forces do not play a significant role in energy consumption. However, the gravity has a great influence in the energy consumption, given the higher cutting resistance implied.

Fig. 22 provides the relationship between the energy consumption and the gravity field on both high-vacuum and non-vacuum environments when the cutting displacement is 0.0825 m. The energy consumption can be nearly fitted by a linear line under both conditions. Nevertheless, it can be observed that the energy consumption corresponding to extraterrestrial conditions is a bit higher, as shown in both Figs. 21 and 22.

Fig. 23 presents the energy consumption due to van der Waals forces under different gravity fields. It is shown that the additional resistance decreases with the gravity field because of the decreased initial stress. However, the proportion of energy consumption due to van der Waals force increases rapidly when the gravity field decreases. In fact, when the gravity is 5 g, the proportion is 2.6% which can be ignored, while it is 16.8% which should not be ignored when the gravity is 1/6 g. So in lunar excavation, the van der Waals forces should not be ignored.

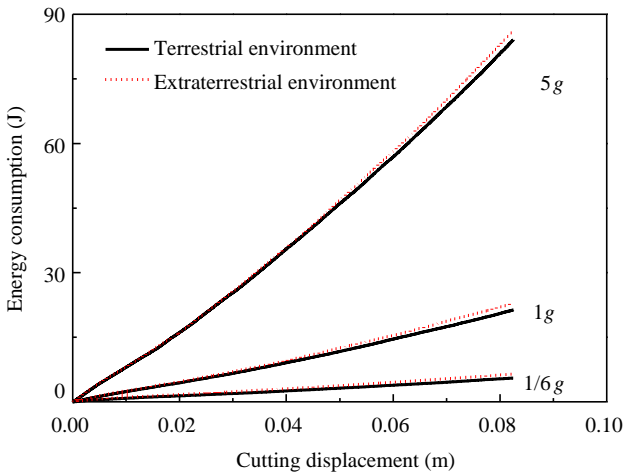

Fig. 21. Evolution of energy consumption under different conditions.

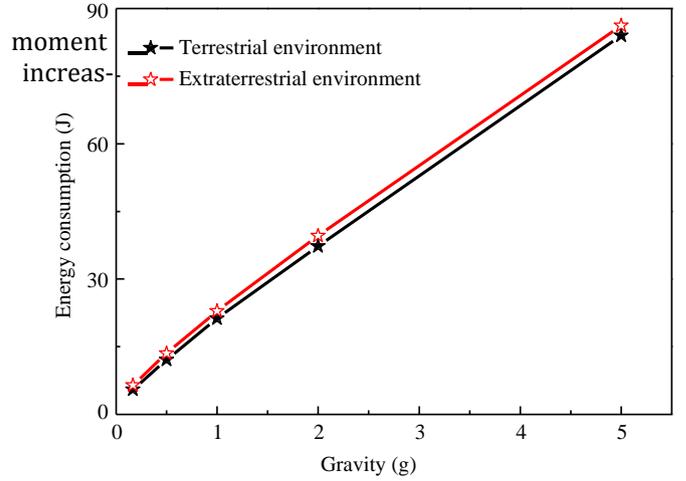

Fig. 22. Relationship between energy consumption and gravity acceleration.

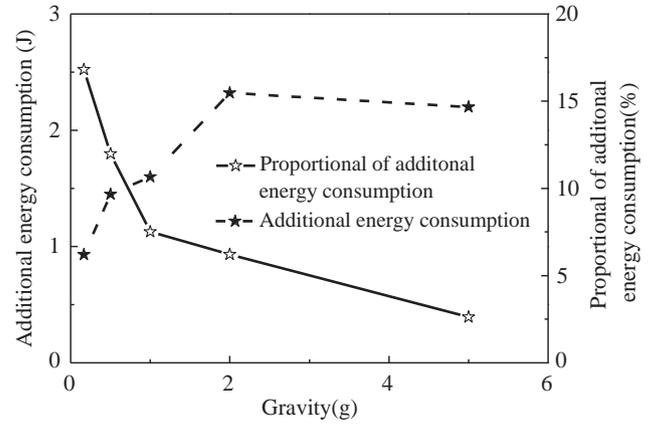

Fig. 23. Additional energy consumption due to van der Waals forces.

6. Concluding remarks

In this paper, Our work are carried out in the following three aspects: (1) a microscopic contact model that can capture the mechanical behavior of lunar regolith is employed; (2) the model has been verified to obtain the mechanical properties of the real lunar regolith in a quasi-static problem; (3) the viscous damping is used in our paper to analyze the soil cutting test, which is a dynamic problem, and the value of the viscous damping was determined by matching the DEM repose angle with the experimental value in slump test. Hence, our simulation in this paper has been validated indirectly and can provide meaningful results about the effects of the gravity field and the van der Waals force due to the unique environment existing on the Moon. The major conclusions are as follows:

1. The gravity field shows significant effects on soil cutting tests. When gravity decreases, the initial stress decreases proportionally, and the affected area and the major slide surface becomes larger. As a result, the cutting resistance and the energy consumption decrease linearly. In addition, the soil heap structure becomes more precipitous and the moment arm becomes shorter in low gravity fields, which leads to a fast increasing rate of bending moment in low gravity fields.
2. The affected area and slide surface becomes smaller if we take the van der Waals forces into consideration, but the cutting resistance, the energy consumption and the bending moment increase significantly because of the raised soil strength. Especially in low gravity fields, the additional cutting resistance, moment and energy consumption takes a large percentage of their total values.
3. In lunar excavation, the maximum frictional force provided by the excavator deceases proportionally with the gravity field, while the cutting resistances decrease slowly because of the larger affected area and the van der Waals forces. Therefore, the same excavator that works efficiently on the Earth may not work properly on the Moon.

Because of the unique lunar environment, the excavator should be well designed and work in a higher efficacy to accomplish a excavation mission on the Moon. More researches on the effects of lunar environment are necessary.

Acknowledgements

The research is funded by the National Natural Science Foundation of China with Grant Nos. 51179128 and 51579178, and State Key Lab. of Disaster Reduction in Civil Engineering (No. 300 SLDRCE14-A-04), all of which are sincerely appreciated. Support of the EU through grants RFPVII - IRSES-GA-2011-294976 and H2020 RISE-GA-645665 is also duly acknowledged. The authors would thank the first author's PhD student Zhifu Shen and former MSc student Yongsheng Dai for their assistance in this numerical simulation.

References

Heiken, G.H., Vaniman, D.T., French, B.M., 1991. Lunar Sourcebook: A User's Guide to the Moon. Cambridge University Press, London.
Ouyang, Z.Y., 2005. Introduction to Lunar Science. China Astronautic Publishing House, Beijing.
Godwin, R.J., Spoor, G., 1977. Soil failure with narrow tines. J. Agric. Eng. Res. 22 (3), 213–228.
Patel, B.P., Prajapati, J.M., 2011. Soil-tool interaction as a review for digging operation of mini hydraulic excavator. Int. J. Eng. Sci. Technol. 3 (2), 894–901.
Zeng, X., Burnoski, L., Agui, J.H., Wilkinson, A., 2007. Calculation of excavation force for ISRU on lunar surface. In: 45th AIAA Aerospace Science Meeting and Exhibit.
McKyes, E., Ali, O.S., 1977. The cutting of soil by narrow blades. J. Terrramech. 14 (2), 43–58.
Hettiaratchi, D.R.P., Reece, A.R., 1967. Symmetrical three-dimensional soil failure. J. Terrramech. 4 (3), 45–67.
Blouin, S., Hemami, A., Lipsett, M., 2001. Review of resistive force models for earthmoving processes. J. Aerosp. Eng. 14 (3), 102–111.
Kuczewski, J., Piotrowska, E., 1998. An improved model for forces on narrow soil cutting tines. Soil Tillage Res. 46 (3), 231–239.
Willman, B.M., Boles, W.W., 1995. Soil-tool interaction theories as they apply to lunar soil simulant. J. Aerosp. Eng. 8 (2), 88–99.
Grisso, R.D., Perumpral, J.V., 1985. Review of models for predicting performance of narrow tillage tool. Trans. ASAE 28 (4), 1062–1067.
Kobayashi, T., Ochiai, H., Fukagawa, R., Aoki, S., Tamoi, K., 2006. A proposal for estimating strength parameters of lunar surface from soil cutting resistances. Earth Space 2 (3).
Boles, W.W., Scott, W.D., Connolly, J.F., 1997. Excavation forces in reduced gravity environment. J. Aerosp. Eng. 10 (2), 99–103.
Boles, W.W., Connolly, J.F., 1996. Lunar excavating research. Eng. Constr. Operat. Space V. ASCE, pp. 699–705.
Green, A., Zacny, K., Pestana, J., Lieu, D., Mueller, R., 2012. Investigating the effects of percussion on excavation forces. J. Aerosp. Eng. 26 (1), 87–96.
Onwualu, A.P., Watts, K.C., 1998. Draught and vertical forces obtained from dynamic soil cutting by plane tillage tools. Soil Tillage Res. 48 (4), 239–253.
King, R.H., Van Susante, P., Gefreh, M.A., 2011. Analytical models and laboratory measurements of the soil–tool interaction force to push a narrow tool through JSC-1A lunar simulant and Ottawa sand at different cutting depths. J. Terrramech. 48 (1), 85–95.
King, R.H., Brewer, A.T., 2012. Laboratory-scale distributed pressure measurements of blade interaction with JSC-1A lunar simulant. J. Aerosp. Eng. 26 (1), 105–116.
Johnson, L.L., King, R.H., 2010. Measurement of force to excavate extraterrestrial regolith with a small bucket-wheel device. J. Terrramech. 47 (2), 87–95.
Iai, M., Gertsch, L., 2012. Excavation of lunar regolith with large grains by rippers for improved excavation efficiency. J. Aerosp. Eng. 26 (1), 97–104.
Agui, J.H., Bucek, M., DeGennaro, A., Wilkinson, R.A., Zeng, X., 2012. Lunar excavation experiments in simulant soil test beds: revisiting the surveyor geotechnical data. J. Aerosp. Eng. 26 (1), 117–133.
Perko, H.A., Nelson, J.D., Sadeh, W.Z., 2001. Surface cleanliness effect on lunar soil shear strength. J. Geotech. Geoenviron. Eng. 127 (4), 371–383.
Bromwell, L.G., 1966. The friction of quartz in high vacuum Res. in Earth Physics Phase Rep. No. 7, R66-18. Dept. of Civ. Eng., Massachusetts Institute of Technology, Cambridge.
Nelson, J.D., 1967. Environmental Effects on Engineering Properties of Simulated Lunar Soils PhD thesis. Illinois Institute of Technology, Chicago.
Johnson, B.V., Roepke, W.W., Strebig, K.C., 1973. Shear testing of simulated lunar soil in ultrahigh vacuum Report No. NASA-CR-09-040-001. U.S. Bureau of Mines.
Johnson, S.W., Lee, D.G., Pyrz, A.P., Thompson, J.E., 1970. Simulating the effects of gravitational field and atmosphere on behavior of granular media. J. Spacecr. Rockets 7 (11), 1311.
Abo-Elnor, M., Hamilton, R., Boyle, J.T., 2003. 3D Dynamic analysis of soil-tool interaction using the finite element method. J. Terrramech. 40 (1), 51–62.
Abo-Elnor, M., Hamilton, R., Boyle, J.T., 2004. Simulation of soil–blade interaction for sandy soil using advanced 3D finite element analysis. Soil Tillage Res. 75 (1), 61–73.
Zhong, J., Zhang, X., Jiang, J., 2010. Study of soil-blade interaction based on finite element method and classical theory. In: Advanced Technol-ogy of Design and Manufacture (ATDM 2010), International Conference on IET, pp. 138–142.

Cundall, P.A., Strack, O.D.L., 1979. A Discrete numerical model for granular assemblies. Géotechnique 29 (1), 47–65.

Mak, J., Chen, Y., Sadek, M.A., 2012. Determining parameters of a discrete element model for soil–tool interaction. Soil Tillage Res. 118 (5), 117–122.

Chen, Y., Munkholm, L.J., Nyord, T., 2013. A discrete element model for soil–sweep interaction in three different soils. Soil Tillage Res. 126 (1), 34–41.

Coetzee, C.J., Els, D.N.J., 2009. Calibration of granular material parameters for DEM modelling and numerical verification by blade–granular material interaction. J. Terrramech. 46 (1), 15–26.

Asaf, Z., Rubinstein, D., Shmulevich, I., 2007. Determination of discrete element model parameters required for soil tillage. Soil Tillage Res. 92 (1–2), 227–242.

Obermayr, M., Klaus, Dressler, Vrettos, C., Eberhard, P., 2011. Prediction of draft forces in cohesionless soil with the Discrete Element Method. J. Terrramech. 48 (5), 347–358.

Coetzee, C.J., Els, D., 2009. The numerical modelling of excavator bucket filling using DEM. J. Terrramech. 46 (5), 217–227.

Shmulevich, I., 2010. State of the art modeling of soil-tillage interaction using discrete element method. Soil Tillage Res. 111 (1), 41–53.

Shmulevich, I., Asaf, Z., Rubinstein, D., 2007. Interaction between soil and a wide cutting blade using the discrete element method. Soil Tillage Res. 97 (1), 37–50.

Nakashima, H., Shioji, Y., Kobayashi, T., Aoki, S., Shimizu, H., Miyasaka, J., et al, 2011. Determining the angle of repose of sand under low-gravity conditions using discrete element method. J. Terrramech. 48 (1), 17–26.

Nakashima, H., Shioji, Y., Tateyama, K., Aoki, S., Kanamori, H., Yokoyama, T., 2008. Specific cutting resistance of lunar regolith simulant under low gravity conditions. J. Space Eng. 1 (1), 58–68.

Bui, H.H., Kobayashi, T., Fukagawa, R., Wells, C.J., 2009. Numerical and experimental studies of gravity effect on the mechanism of lunar excavations. J. Terrramech. 46 (3), 115–124.

Jiang, M.J., Shen, Z.F., Thornton, C., 2013. Microscopic contact model of lunar regolith for high efficiency discrete element analyses. Comput. Geotech. 54 (10), 104–116.

Jiang, M.J., Liu, F., Shen, Z.F., Zheng, M., 2014a. Distinct element simulation of lugged wheel performance under extraterrestrial environmental effects. Acta Astronaut. 99 (2), 37–51.

Jiang, M.J., Wang, X.X., Zheng, M., Dai, Y.S., 2014b. Interaction between lugged wheel of lunar rover and lunar soil by DEM with a new contact model. Earth and Space 2012 (ASCE), 155–164.

Jiang, M.J., Xi, B.L., Shen, Z.F., Dai, Y.S., 2015a. DEM analyses of horizontal pushing resistance under different gravity fields. Chin. J. Geotech. Eng. 37 (7), 1300–1306.

Jiang, M.J., Li, L.Q., Sun, Y.G., 2015b. Properties of TJ-1 lunar soil simulant. J. Aerosp. Eng. 25 (3), 463–469.

Dai, Y.S., 2013. Experimental and numerical investigation on the wheel-soil interaction Msc thesis. Tongji University, Shanghai, China (in Chinese).

Jiang, M.J., Li, L.Q., Yang, Q.J., 2013a. Experimental investigation on deformation behavior of TJ-1 lunar soil simulant subjected to principal stress rotation. Adv. Space Res. 52 (1), 136–146.

Jiang, M.J., Li, L.Q., Liu, F., Wang, C., 2013b. Deformation behavior of TJ-1 lunar soil simulant under principal stress rotation. J. Tongji Univ. 41 (4), 483–489.

Jiang, M.J., Zhang, N., Shen, Z.F., Wu, X., 2013c. CPT-based estimation of bearing and deformation indexes for TJ-1 lunar soil simulant ground. In: Powders and Grains: Proceedings of the 7th International Conference on Micromechanics of Granular Media, Sydney, Australia, pp. 309–312.

Jiang, M.J., Konrad, J.M., Leroueil, S., 2003. An efficient technique for generating homogeneous specimens for DEM studies. Comput. Geotech. 30 (03), 579–597.